\documentclass[journal=jcisd8,manuscript=article,layout=onecolumn]{achemso}
\setkeys{acs}{articletitle=true, maxauthors=99}
\usepackage[utf8]{inputenc}
\usepackage[T1]{fontenc}
\usepackage{lineno}
\usepackage{chemformula} 
\usepackage{multicol}
\usepackage{longtable}
\usepackage{amsmath, amsfonts}
\usepackage{bm}
\usepackage{hyperref}
\hypersetup{colorlinks=true, allcolors=blue}

\newcommand{\etal}{\textit{et al.}}

\title{\textbf{Improving the Silicon Interactions of GFN-xTB}}

\author{Leonid Komissarov}
\affiliation[CCM]{Center for Molecular Modeling (CMM), Ghent University, Technologiepark-Zwijnaarde 46, B-9052, Ghent, Belgium}

\author{Toon Verstraelen}
\affiliation[CCM]{Center for Molecular Modeling (CMM), Ghent University, Technologiepark-Zwijnaarde 46, B-9052, Ghent, Belgium}
\email{toon.verstraelen@ugent.be}

\abbreviations{}

\begin{document}

\begin{abstract}
A general-purpose Density Functional Tight Binding method, the GFN-xTB model is gaining increased popularity in accurate simulations that are out of scope for conventional \textit{ab initio} formalisms.
We show that in its original GFN1-xTB parametrization, organosilicon compounds are described poorly.
This issue is addressed by re-fitting the model's silicon parameters to a data set of ten thousand reference compounds,
geometry-optimized with the revPBE functional.
The resulting GFN1-xTB-Si parametrization shows improved accuracy in the prediction of system energies, nuclear forces and geometries
and should be considered for all applications of the GFN-xTB Hamiltonian to systems that contain silicon.
\end{abstract}

\newpage

\section{Background \& Summary}
Silicon is the second most abundant element on earth.\cite{crc}
Its physical properties make it a crucial building block for applications in electronics\cite{si-appl1, si-appl2, si-appl3} and
materials science.\cite{si-appl4, si-appl5, si-appl6}
More recently organosilicon compounds have become of increasing interest to the field of organic synthesis in the role of (selective) intermediates\cite{si-appl7, si-appl8, si-appl9}
or catalysts.\cite{si-appl10, si-appl11}
One way scientists can study any of the aforementioned applications is through computational simulations of appropriate systems on modern computer hardware.
Compared to experimental laboratory work, this domain of computational chemistry offers a high amount of flexibility when it comes to throughput, man-hours and overall cost.
A myriad of models can be employed to simulate a system of interest -- too many to give a a comprehensive overview of all of them here
(we point the reader to Refs.\ \citenum{cramer2004essentials, frenkel-smit} for a comprehensive introduction to the topic).
Instead, we roughly classify two fundamental model types: \textit{ab initio} and \textit{empirical}.
The former describes a model that is derived from the fundamental laws of physics and, without introducing approximations fitted to (experimental) reference data.
Examples of \textit{ab initio} models are the Hartree-Fock (HF) formalism, configuration interaction (CI) methods and density functional theory (DFT) with non-empirical functionals like PBE \cite{perdew_generalized_1996}.
Explicit treatment of the electronic structure means that \textit{ab initio} computations can be highly accurate, but computationally slow and only limited to small systems of roughly tens to hundreds of atoms.
Empirical models in contrast are fast and can handle system sizes of up to millions of atoms.
This speed-up is achieved through simplifications in the description of interatomic interactions. Examples for empirical models are classical force fields (FF) or machine learning potentials (MLP).
One benefit of empirical models is the ability to fit their parameters to a specific chemical space. This allows for an improvement in prediction accuracy without the need to switch to a computationally more expensive model.
The semi-empirical GFN1-xTB\cite{gfn1xtb} formalism falls under this category.
The model has found a broad application in the modeling of small-to-medium organic molecules,
where it is predominantly used for geometry optimizations  \cite{gfn-appl1, gfn-appl2, gfn-appl3, gfn-appl4}.
Despite its success, we have observed large discrepancies between GFN1-xTB and DFT when comparing relative energies, optimized geometries and nuclear gradients of organosilicon compounds.
This issue is addressed by fitting the silicon parameters of GFN1-xTB to higher-level DFT data.
We describe a reference data set of 10000 organosilicon compounds, followed by the parameter optimization procedure in the Methods Section.
The Results Section discusses shortcomings of the original GFN1-xTB model in more detail and compares it to our newly obtained parameters as well as to similar models from the literature.
In contrast to the original parametrization, our parameters reproduce system energies and geometries more accurately,
without compromise to the prediction accuracy of non-silicon organic compounds, as tested on a subset of the ANI-1x\cite{ani1x-data} data.

\section{Methods}

\subsection*{Organosilicon Reference Data}
Initial structures for our reference data set are taken from the
PubChem library.\cite{pubchem1, pubchem2}
The search query included the following filters:
\begin{itemize}
    \item Heavy atom count between 1 and 15
    \item Compound contains Si, O, C
    \item Covalent unit count of 1
    \item Molecular weight less or equal to 200 g/mol
    \item Total formal charge of 0
\end{itemize}
From the total of 50750 compounds matching the query, a random subset of 10k structures is selected.
The resulting set has a mean heavy atom count of 10.5 (smallest: 3, largest: 13, standard deviation: 1.9).
In addition to the chemical elements defined in the search query, H, N, Cl, S, F, P and Br are included in the set.
Elemental occurrences are presented in Table \ref{tab:occurrences}, with nitrogen being the most, and bromine the least prominent among the additional elements (other than Si, O and C) in the set.
All structures are geometry-optimized with the Amsterdam Density Functional\cite{adf} (ADF) molecular simulation package,
as integrated in the Amsterdam Modeling Suite\cite{ams}.
We use the revPBE functional, \cite{pbe, revpbe} a 'Small' frozen core and the double-zeta polarized (DZP) basis set.
Geometry convergence criteria are left at their default values, namely
0.001 Hartree/Å, 0.00001 Hartree/Atom and 0.1 Å for atomic gradients, energy and atomic displacements respectively.
A Quasi-Newton optimizer\cite{quasi-newton} in the delocalized coordinates space is used for the optimizations.
Distributions of all the convergence criteria at each structure's last optimization step are provided in Figure S1.
Training and validation data sets are constructed from
(1) atomic forces at the initial (un-optimized) geometries, considering Si atoms only,
(2) energy differences between the same compound's initial and optimized geometries,
(3) root-mean-square deviation (RMSD) of the optimized geometries.
Prior to the construction of the data sets, improbable outlier geometries are filtered by excluding 
123 systems with an energy difference larger than \mbox{5 kJ/mol} per atom between optimized and initial geometries.
\begin{table}
\centering
\caption{
    Elemental occurrences in the reference data set.
    Structures containing at least one atom of the listed element are counted in the upper part.
    Structures with the exact number of silicon atoms are counted in the lower part.
    }
\begin{tabular}{lr}
    \textbf{Element}& \textbf{Occurrence}\\
    \hline
C&	10000 \\
O&	10000 \\
Si&	10000 \\
H&	9994 \\
N&	3252 \\
Cl&	573 \\
S&	461 \\
F&	295 \\
P&	85 \\
Br&	27 \\
\hline
1 Si&   9318 \\
2 Si&   613 \\
3 Si&   59 \\
4 Si&   9 \\
5 Si&   1 \\
\label{tab:occurrences}
\end{tabular}
\end{table}

\subsection{Optimization of the Silicon GFN-xTB Parameters}
Developed by Grimme \etal, the GFN-xTB (also GFN1-xTB) model is a semiempirical method for the computation of a chemical system's Hamiltonian.\cite{gfn1xtb}
The method follows a Density Functional Tight Binding (DFTB) approximation, which describes the electronic energy $E_\mathrm{el}$ of a molecule as a functional of its (valence) electron density $\rho(\bm{r}) = \rho_0(\bm{r}) + \delta\rho(\bm{r})$, where $\bm{r}$ is a spatial coordinate.\cite{dftb_1, sqm_review}
The reference density $\rho_0$ is typically a superposition of individual atomic contributions, whereas $\delta \rho$ is the consequence chemical bonding and is assumed to be relatively small.
In the DFTB formalism, the energy is approximated by a Taylor series up to the third order in $\delta \rho$ (corresponding to DFTB3), such that 
\begin{equation}
    \label{eq:dftbE}
    E_\mathrm{el}[\rho] = E^0[\rho_0] + E^1[\rho_0, (\delta\rho)^1] + E^2[\rho_0, (\delta\rho)^2] + E^3[\rho_0, (\delta\rho)^3].
\end{equation}
The total GFN1-xTB energy is divided into electronic (el), repulsive (rep), dispersion (disp) and halogen-bonding (xb) interactions and can be written as
\begin{equation}
    \label{eq:xtbE}
    E = E_\mathrm{el} + E_\mathrm{rep} + E_\mathrm{disp} + E_\mathrm{xb}.
\end{equation}
Reference \citenum{gfn1xtb} describes the GFN-xTB Hamiltonian in more detail.

The parameter optimization is performed with the ParAMS parameter fitting package.\cite{params}
Starting from Grimme's original parametrization\cite{gfn1xtb}, we optimize all 17 silicon parameters from the electronic and repulsive terms,
plus one parameter specific to the Si-O atom-pair for a total of 18 parameters.
A summary of all optimized parameters, and their values in the original parametrization are provided in Table \ref{tab:params}
along with upper and lower parameter ranges that were used during the optimization.
Training and validation sets are created by randomly splitting the complete data set into relative sizes of 80 and 20 percent respectively (splitting is based on unique structures, not their properties).
Within the training set, each entry of relative energies, atomic forces and RMSD is assigned a weight of 2.4, 28.0 and 1.0 respectively. 
The weights were determined by minimizing the standard deviation of all per-entry contributions to the overall root-mean-square error (RMSE), as calculated with the initial GFN1-xTB parameters.
Covariance matrix adaptation evolution strategy \mbox{(CMA-ES)}\cite{cma1, cma2}, a gradient-free, population-based optimization algorithm
is used to optimize the GFN-xTB parameters.
Population size and initial sampling width $\sigma$ are set to 12 and 0.2 respectively.
The optimization is set up to run for a maximum of 72 hours.
To prevent a waste of computation time when stuck in local minima, an early stopping algorithm is set up to abort the optimization if there is no improvement
in the training set loss after 1000 evaluations.
At every optimization step, only a batch of 800 randomly selected jobs is computed to speed up convergence.
To compensate for both, the noise introduced through the aforementioned batching and the pseudo-random sampling of CMA-ES,
8 independent optimizations are performed.
The best parameter set is selected based the lowest training set loss function value.

\begin{table}
\centering
\caption{
    List of all optimized parameter names (Si only where applicable, atomic subscript dropped),
    their initial values and the equation number that lists the parameter as described in Ref.\ \citenum{gfn1xtb},
    followed by the optimization bounds used in this work and the optimized values.
    }
\begin{tabular}{lrrrr}
    \textbf{Parameter}& \textbf{Equation}& \textbf{Original Value}& \textbf{Optimization Bounds} &\textbf{Optimized Value}\\
    \hline
    
$\eta$              &   5&    +0.438&   (+0.100,  +3.500)&    +2.251\\
$\Gamma$            &   9&    +1.500&   (+1.000,  +2.000)&    +1.125\\
$\alpha$            &  13&    +0.948&   (+0.500,  +1.800)&    +1.036\\
$Z$                 &  13&   +16.898&   (+8.000, +25.000)&   +21.357\\
$EN$                &  10&    +1.900&   (+1.710,  +2.090)&    +2.089\\
$K_\mathrm{SiO}$    &  10&    +1.000&   (+0.900,  +1.100)&    +0.969\\
\\
\textit{3s level}& & & \\
\hline
$k^\mathrm{poly}_l$ &  11&   -14.202&  (-24.000, +15.000)&   +14.825\\
$\kappa^l$          &   5&    +0.000&  (-10.000, +10.000)&    -4.972\\
$H^l$               &  12&   -14.506&  (-30.000,  -2.000)&   -20.800\\
$\zeta_l$           &   7&    +1.522&   (+0.400,  +4.000)&    +2.337\\
\\
\textit{3p level}& & & \\
\hline
$k^\mathrm{poly}_l$ &  11&    -3.893&  (-10.000, +30.000)&    -8.629\\
$\kappa^l$          &   5&    -5.926&  (-10.926,  -0.926)&    -6.046\\
$H^l$               &  12&    -7.557&  (-17.557,  +2.443)&    -3.526\\
$\zeta_l$           &   7&    +1.609&   (+0.400,  +4.000)&    +1.576\\
\\
\textit{3d level}& & & \\
\hline
$k^\mathrm{poly}_l$ &  11&   +25.499&  (-30.000, +44.000)&   +36.572\\
$\kappa^l$          &   5&    +0.000&  (-10.000, +10.000)&    +6.072\\
$H^l$               &  12&    -2.508&  (-12.508,  +7.492)&    -3.321\\
$\zeta_l$           &   7&    +1.169&   (+0.400,  +4.000)&    +2.474\\

\label{tab:params}
\end{tabular}
\end{table}

\section{Results and Discussion}

\subsection*{Performance of Models from the Literature}
\begin{figure}
\centering
\includegraphics[width=0.4\linewidth]{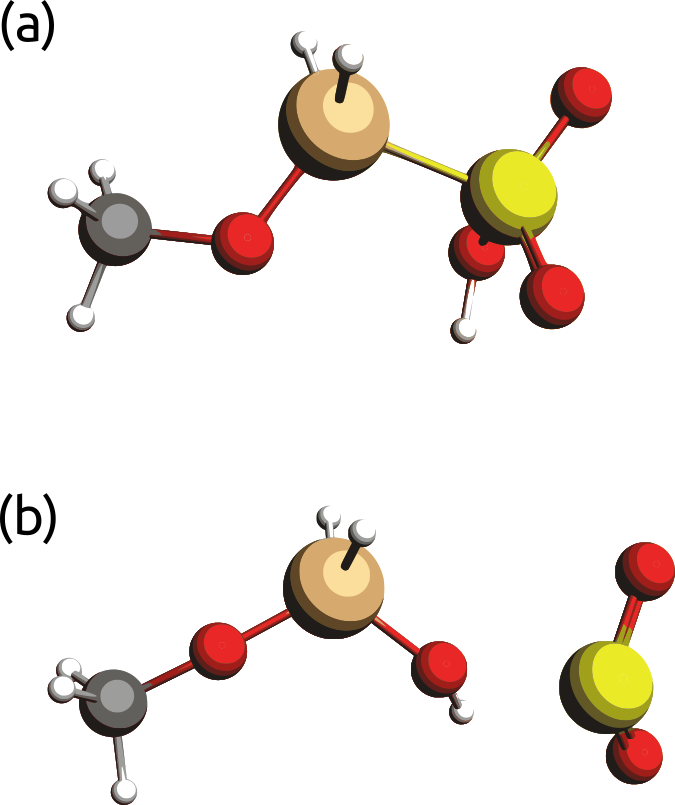}
\caption{
    Optimized geometries of sulfosilyloxymethane (PubChem ID 154240983).
    Structures optimized with (a) revPBE and (b) GFN1-xTB.
    The latter shows a re-arrangement and Si-O-R angles close to 180 degrees.
    Hydrogen, carbon, oxygen, silicon and sulfur depicted in white, grey, red, beige and yellow respectively.
    }
\label{fig:xtb_bloopers}
\end{figure}

To serve as a baseline, the performance of three DFTB parametrizations from the literature is compared on our validation set.
The first three rows of Figure \ref{fig:x0} compare three general-purpose models, namely GFN1-xTB\cite{gfn1xtb} (the initial parameter set used for our optimizations),
GFN2-xTB\cite{gfn2xtb} and the QUASINANO2015\cite{quasinano2015} set.
Overall, GFN2-xTB (Fig. \ref{fig:x0} second row) is the most accurate of the three, with the lowest errors in the energy differences and the RMSD of geometry-optimized structures.
Both GFN models predict atomic forces with the same accuracy (Fig. \ref{fig:x0} middle column, first and second rows).
Although the QUASINANO2015 parameters are the least accurate of the three when comparing relative energies and atomic forces, distributions of the
RMSD are better than for GFN1-xTB.
We found that in most cases, the poor RMSD of GFN1-xTB can be traced back to unrealistic Si-O-R bond angles.
This problem is visualized in Figure \ref{fig:angles}, showing the distributions of all Si-O-C angles for each of the aforementioned models.
Note how DFT predicts an average angle of 121 degrees (Fig. \ref{fig:angles}a),
while most angles computed with GFN1-xTB are almost colinear with an angle close to 180 degrees (Fig. \ref{fig:angles}b).
This issue is not observed for the GFN2-xTB and QUASINANO2015 models (Fig. \ref{fig:angles}c,d).
In more extreme cases geometry optimizations with GFN1-xTB resulted in rearrangements and bond dissociations.
One example is presented in Figure \ref{fig:xtb_bloopers}, showing the optimized structure of sulfosilyloxymethane, as computed with revPBE and GFN1-xTB.

\subsection*{Optimized Silicon Parameters for GFN1-xTB}
\begin{figure}
\centering
\includegraphics[width=0.85\linewidth]{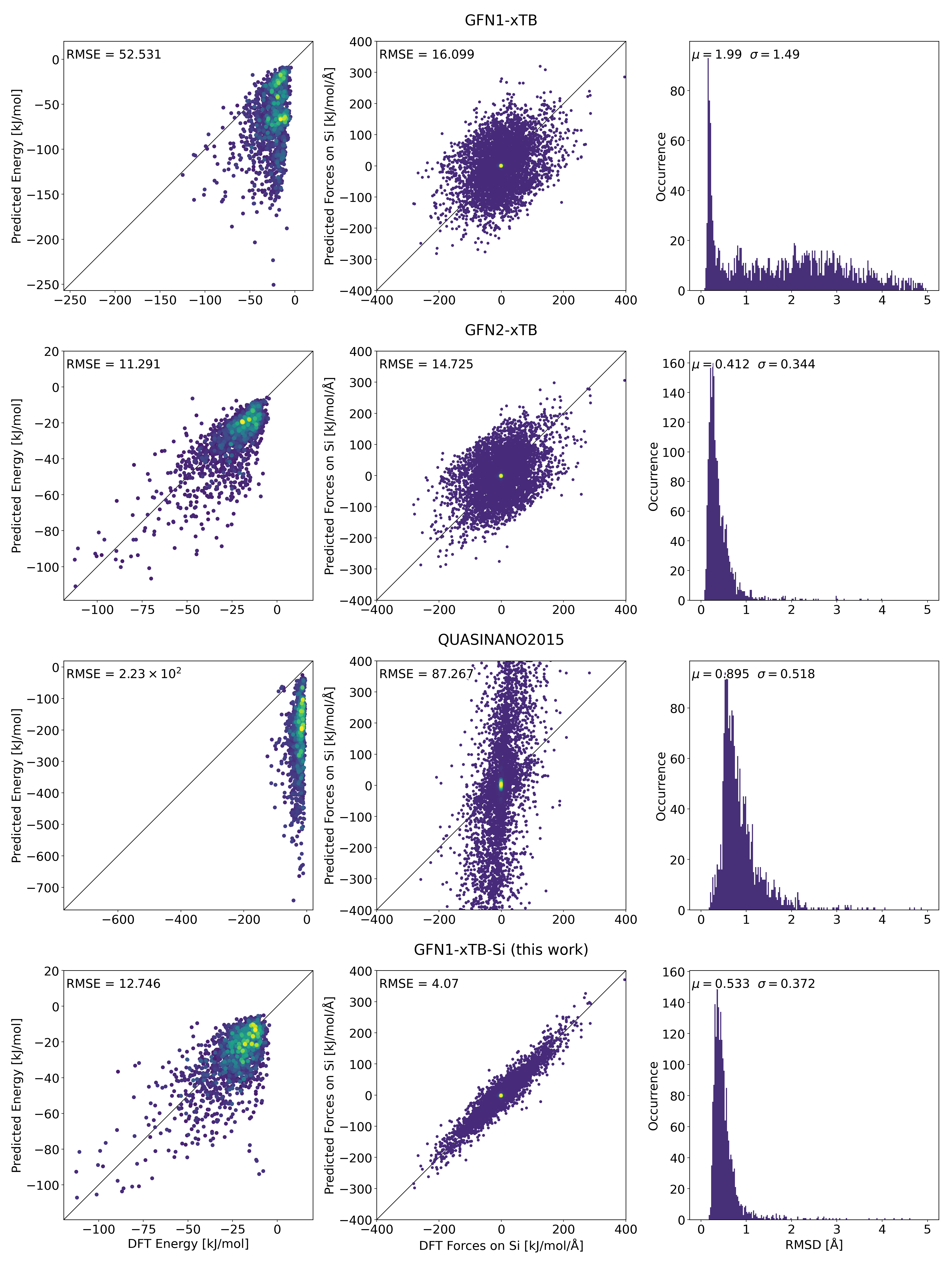}
\caption{
    Validation set performance of various Density Functional Tight Binding models.
    Comparing GFN1-xTB\cite{gfn1xtb}, GFN2-xTB\cite{gfn2xtb}, QUASINANO2015\cite{quasinano2015} and this GFN1-xTB-Si (this work)
    from top to bottom.
    Columns, from left to right depict energy differences, force components on the Si atoms, and RMSD of atomic positions (as described in the Methods Section).
    X and Y values in the first two columns are reference properties and their DFTB predictions respectively.
    Areas of lower point densities are depicted in dark blue; higher densities in bright green.
    Root-mean-square error (RMSE) printed in the same units as the axes.
    Histograms in the right column show the RMSD between geometry-optimized reference and DFTB structures.
    Mean $\mu$ and standard deviation $\sigma$ printed in ångström.
    }
\label{fig:x0}
\end{figure}

\begin{figure}
\centering
\includegraphics[height=0.9\textheight]{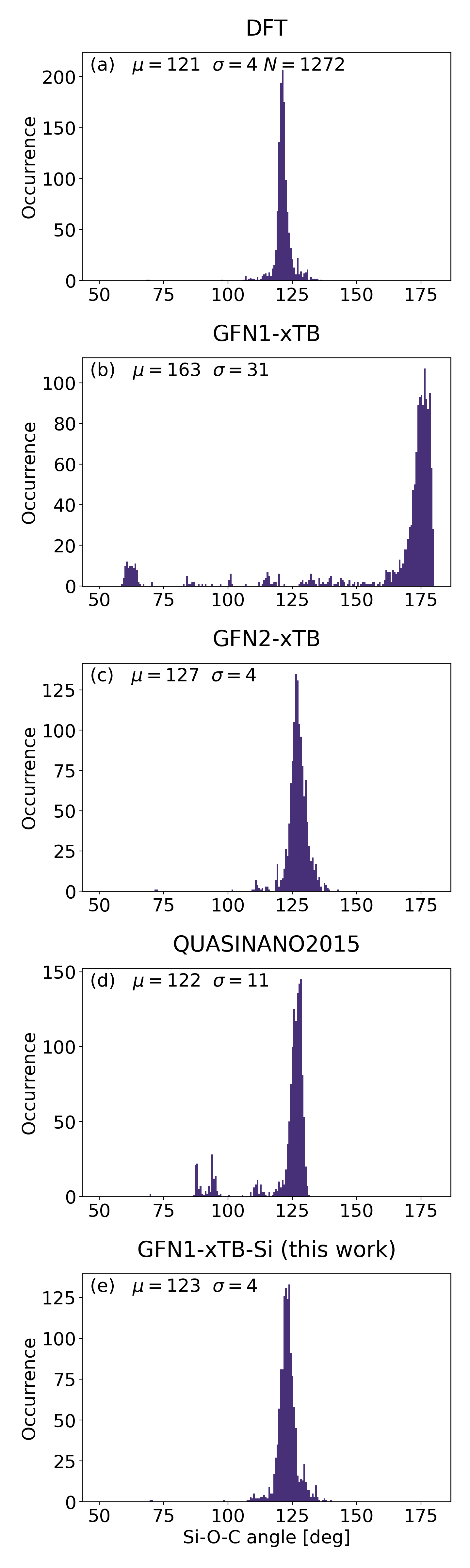}
\caption{
    Si-O-C angle distributions of all optimized structures in the validation set. 
    Comparing the revPBE\cite{revpbe}, GFN1-xTB\cite{gfn1xtb}, GFN2-xTB\cite{gfn2xtb}, QUASINANO2015\cite{quasinano2015} and this GFN1-xTB-Si (this work) in a,b,c,d,e respectively.
    Printed text shows mean ($\mu$), standard deviation ($\sigma$) and the total number of points ($N$).
    }
\label{fig:angles}
\end{figure}

We report all optimized parameter values in Table \ref{tab:params}.
Following the previous section, we compare our new set of parameters, which we refer to as GFN1-xTB-Si,
in the last row of Figure \ref{fig:x0} and Figure \ref{fig:angles}e.
Substantial improvements can be observed in all three of the fitted properties,
bringing GFN1-xTB-Si to an accuracy level that is comparable to GFN2-xTB.
At the same time, calculations with the GFN1-xTB Hamiltonian have shown to be roughly 35\% faster when compared to the GFN2-xTB model.
This has been tested by randomly selecting a batch of 200 geometries from our data set and measuring the time it
took both models to calculate energies and atomic forces for all structures.
The set-up has been repeated 100 times to produce average timings.

A sanity check is performed by calculating the atomic gradients on a separate test set.
We use the ANI-1x reference data by Smith \etal\cite{ani1x-data} for this purpose,
which includes atomic gradients calculated with the $\omega B97X$\cite{wb97x} functional.
The test set is constructed from one randomly selected conformation for each of the 3113 unique configurations in the ANI-1x data.
Atomic gradients are computed with the GFN1-xTB and GFN1-xTB-Si parameters.
As expected, both parametrizations predict the same forces for the test set, since the ANI-1x data does not include any silicon atoms.
A correlation plot between the $\omega B97X$ and GFN1-xTB forces is presented in Figure S2.
For the ANI-1x set, the RMSE of the atomic gradients is roughly \mbox{28.5 kJ/mol/Å}.
This leads us to the conclusion that the GFN1-xTB-Si parametrization can be used without any compromises when computing energies, gradients and geometries of isolated organic compounds.

We would like to stress that this work introduces a parameter set that has been fitted to geometries, nuclear gradients and energies, and as such should only be used for applications related to these properties.
This forms a contrast to the original claim of the GFN model -- the ability to represent
\textbf{G}eometries, \textbf{F}requencies and \textbf{N}on-covalent interactions accurately,
as the latter two properties have neither been investigated nor fitted here.
Although this calls for a future investigation,
we argue that a first focus on corrected geometries, forces and energies is most logical since
(1) optimized geometries are the most popular prediction target when using GFN1-xTB\cite{gfn-appl1, gfn-appl2, gfn-appl3, gfn-appl4} and
(2) we expect that predictions of frequencies and non-covalent interactions on systems with poorly-described geometries and energies will be of limited interest.
Furthermore, the reduced error on the silicon forces for slightly distorted geometries suggests that the Hessian of the potential energy improved as well, which should be beneficial for the prediction of vibrational frequencies.

\section{Data and Software Availability}
GFN1-xTB-Si parameters (in the AMS format), reference data, including training and validation sets, and files needed to run the parameter optimization scheme are available at DOI:
\href{https://doi.org/10.24435/materialscloud:14-4m}{10.24435/materialscloud:14-4m}
or the Materials Cloud Archive
\href{https://archive.materialscloud.org/record/2021.152}{record 2021.152}.
See the Supporting Information or data repository for a description of the reference data formats.
For the ANI-1x test set, please refer to Ref.\ \citenum{ani1x-data}.
The Amsterdam Modeling Suite\cite{ams} (v. 2020.203),
which includes the ParAMS package\cite{params} (v.\ 0.5.1),
is a commercial software, for which a free trial may be requested at \href{www.scm.com}{www.scm.com}. 

\section{Acknowledgements}
This project has received funding from the European Union's Horizon
2020 research and innovation programme under grant agreement No 814143.
T.V. acknowledges funding of the research board of Ghent University.
The computational resources (Stevin Supercomputer Infrastructure) and services used in this work were provided by the VSC (Flemish Supercomputer Center), funded by Ghent University, FWO and the Flemish Government – department EWI.

\section*{Author contributions statement}
L.K. designed and performed the study. Both authors wrote the manuscript. T.V. oversaw the project

\section*{Competing interests}
The authors declare no competing interests.

\bibliography{refs}

\begin{tocentry}
\includegraphics[width=3.25in]{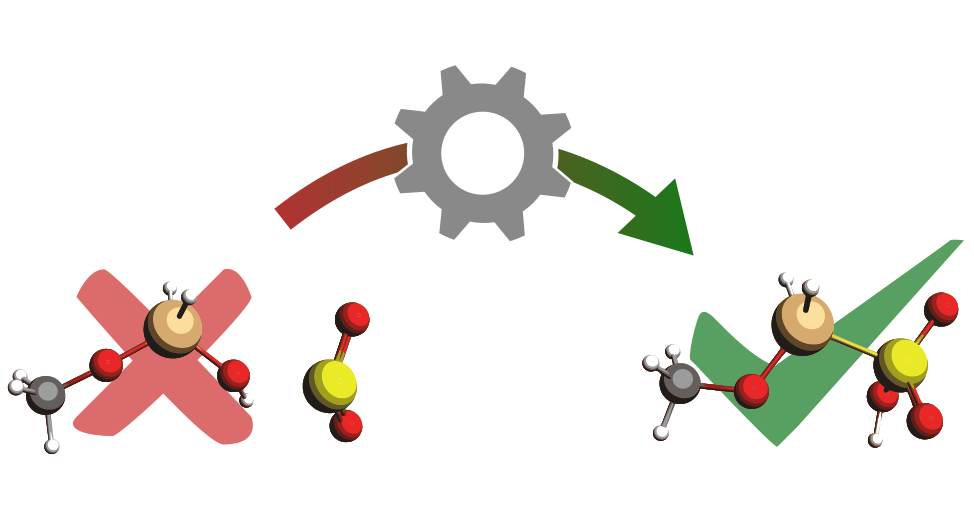}
\end{tocentry}

\end{document}


\flushbottom
\maketitle

\section*{S1. Extra display items}

\begin{figure}
\centering
\includegraphics[width=\linewidth]{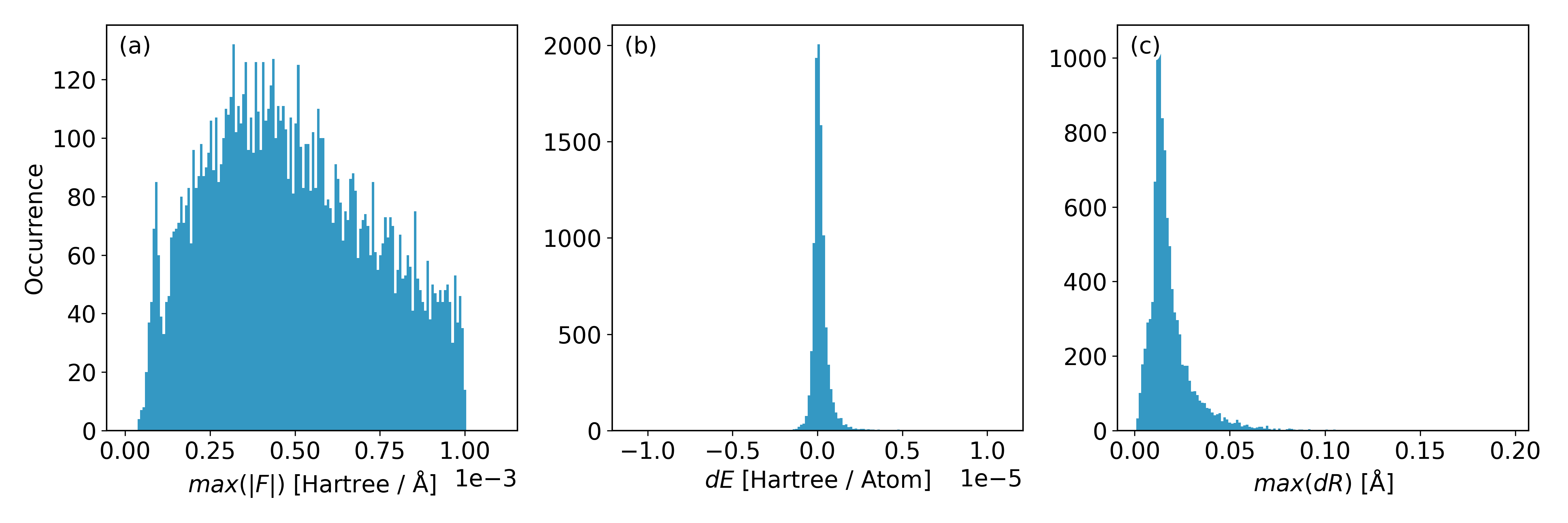}
\caption{
    Distribution of convergence criteria at the last
    optimization step for all calculated systems in the reference data set.
    Showing (a) the highest absolute component of all nuclear gradients,
    (b) change in system energy and (c) highest relative atomic displacement.
    }
\label{fig:convergence}
\end{figure}

\begin{figure}
\centering
\includegraphics[width=0.5\linewidth]{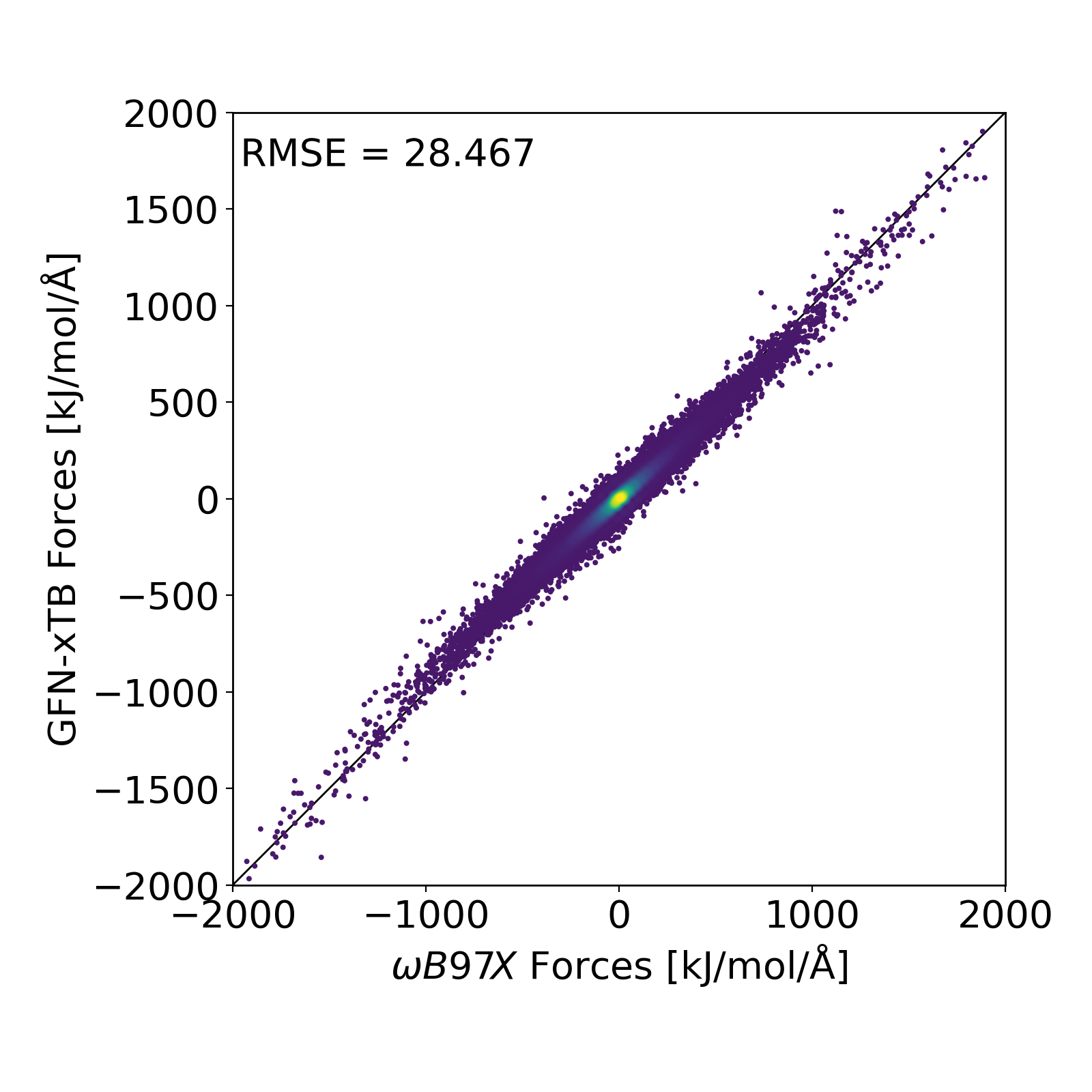}
\caption{
    Correlation plot of atomic forces, as calculated with the test set generated from ANI-1x\cite{ani1x-data} data ($\omega WB97X$\cite{wb97x} reference, x-axis).
    As no silicon is present in this set, both, the GFN1-xTB and GFN1-xTB-Si parametrizations predict the same forces (y-axis).
    }
\label{fig:convergence}
\end{figure}

\begin{table}
\centering
\caption{
    Format specification for the reference data repository.
    Each individual geometry optimization trajectory is stored in a NumPy .npz file with available keys listed below.
    Variables $N$ and $R$ denote the number of geometry optimization steps
    and the system size respectively.
    }
\begin{tabular}{lrrr}
    \textbf{Data}& \textbf{Unit}& \textbf{Key}& \textbf{Array Shape} \\
    \hline
    Atomic Numbers& -& \verb|numbers|& $(R,)$ \\    
    Atomic Coordinates& Å& \verb|xyz|& $(N, R, 3)$ \\
    Energy& hartree& \verb|energy|& $(N,)$ \\
    Nuclear Gradients& hartree/bohr& \verb|gradients|& $(N, R, ,3)$ \\
    
\label{tab:keys}
\end{tabular}
\end{table}

\newpage
\bibliography{refs}